\documentclass[a4paper]{jpconf}
\usepackage{amsfonts,amssymb,
amscd,amsmath,latexsym,amsbsy,bm}
\usepackage{graphicx}
\begin{document}
\title{New Pentagon Identities Revisited}

\author{Shahriyar Jafarzade}
\address{Department of Physics, Mimar Sinan Fine Arts University, Bomonti 34380, Istanbul, Turkey}

\ead{shahriyar.jzade@gmail.com}

\begin{abstract}
We present a new solution to the pentagon identity in terms of gamma function. We obtain this solution by taking the gamma function limit  from the pentagon identity related to the three-dimesional index. This limit corresponds to the identification of the sphere partition function of dual theories and being equivalent to the star-triangle relation in statistical mechanics which corresponds to the ``strongly coupled'' regime of the Faddeev-Volkov model.

\end{abstract}

\section{Introduction}
%%%%%%%%%%%%%%%%%%%%%%%%%%%%%%%%%%%
Following identity is called the pentagon identity \cite{Faddeev:1993rs} for non-commutative operators $xy=qyx$
\begin{align}\label{PI-Operator}
    l(y)l(x)=l(x)l(-xy)l(y),
\end{align}
where $ l(x)=\prod_{i=1}^{\infty}(1-xq^i)$ and $|q|<1$ for complex variable $x$.

This is the quantum generalization of the classical pentagon identity  
\begin{align}
    L(x)+L(y)-L(xy)=L\Big(\frac{x-xy}{1-xy}\Big)+L\Big(\frac{y-xy}{1-xy}\Big),
\end{align}
where Rogers' dilogarithm $L(x)$ is defined as 
\begin{align}
L(x):=-\int_0^x\frac{\log(1-z)}{z}dz+\log(1-x)\log(x)/2.
\end{align}

Notice that, pentagon identity for the quantum dilogarithm has been interpreted as the restricted star-triangle relation in statistical mechanics \cite{Faddeev:1993rs} and Pachner-moves \cite{Hikami:2001en}.

Consider the following five-term relation \footnote{ Here we define pentagon identity in the integral form.} 
\begin{align} \label{PI}
\int d[\sigma] \mathcal{B}\mathcal{B}\mathcal{B}=\mathcal{B}\mathcal{B}.
\end{align}
Recently, one solution of this identity is observed  in terms of hyperbolic gamma function and plays very important role in shaped triangulation \cite{Kashaev:2012cz}. \footnote{For the case of ideal triangulation, see \cite{Dimofte:2011py,Garoufalidis:2017xah}}

Integral identity for this case reads
\begin{align}\label{Pentagon-Hyperbolic}
 \int_{-i\infty}^{i\infty}\frac{du}{i\sqrt{\omega_1\omega_2}} \prod_{i=1}^{3} \mathcal{B}\big(a_i+u,b_i-u\big)=\mathcal{B}\big(a_1+b_2,a_3+b_1\big)\mathcal{B}\big(a_2+b_1,a_3+b_2\big),
\end{align}
where $\mathcal{B}(x,y)$ is defined as 
\begin{align}
    \mathcal{B}(x,y):=\frac{\gamma^{(2)}(x;\omega_1,\omega_2)\gamma^{(2)}(y;\omega_1,\omega_2)}{\gamma^{(2)}(x+y;\omega_1,\omega_2)}.
\end{align}
The hyperbolic gamma function is defined as
\begin{equation}\label{Hyp-Gam}
\gamma^{(2)}(u;\omega_1, \omega_2):=e^{-\pi i B_{2,2}(u;\mathbf{\omega})/2} \frac{(e^{2 \pi i u/\omega_1}\tilde{q};\tilde{q})}{(e^{2 \pi i u/\omega_1};q)} \quad \text{with} \quad q:=e^{2 \pi i \omega_1/\omega_2}, \quad \tilde{q}:=e^{-2 \pi i \omega_2/\omega_1} \; ,
\end{equation}
where  $B_{2,2}(u;\mathbf{\omega})$ is the second order Bernoulli polynomial,
\begin{equation}\label{B_{2,2}} B_{2,2}(u;\mathbf{\omega}):=
\frac{u^2}{\omega_1\omega_2} - \frac{u}{\omega_1} -
\frac{u}{\omega_2} + \frac{\omega_1}{6\omega_2} +
\frac{\omega_2}{6\omega_1} + \frac 12.
\end{equation}
Identity (\ref{Pentagon-Hyperbolic}) describe the matching of $S^3_b$ supersymmetric partition functions \cite{Hosomichi:2014hja} of 3-d duality with $U(1)$ gauge theories and six chiral multiplet for electric part, while no gauge degree of freedom and 9 chiral multiplets for magnetic part. \footnote{See \cite{Dimofte:2011py, Gahramanov:2014ona, Bozkurt:2018xno} for pentagon identities from gauge theories.}

Considering the same duality for three dimensional index leads  to the following pentagon identity \cite{Gahramanov:2016wxi} \footnote{
Special case of this identity corresponds to the pentagon identity related to the tetrahedron index \cite{Gahramanov:2013rda}.}
\begin{align}\nonumber \label{Pentagon-Index}
    \sum_{m=-\infty}^{\infty}  \oint \frac{1}{z^{3m}} \frac{dz}{2\pi i z} &\prod_{i=1}^{3} \mathcal{B}\big(a_iz,n_i+m;b_iz^{-1},m_i-m\big)\\
    &\quad=\mathcal{B}\big(a_1b_2,n_1+m_2;a_3b_1,n_3+m_1\big)\mathcal{B}\big(a_2b_1,n_2+m_1;a_3b_2,n_3+m_2\big),
\end{align}
where the corresponding $\mathcal{B}(a,n;b,m)$ is defined as
\begin{align}
    \mathcal{B}(a,n;b,m):=\frac{(q^{1+\frac{n}{2}}a^{-1};q)_{\infty}}{(q^{\frac{n}{2}}a;q)_{\infty}}\frac{(q^{1+\frac{m}{2}}b^{-1};q)_{\infty}}{(q^{\frac{m}{2}}b;q)_{\infty}}\frac{(q^{\frac{n+m}{2}}ab;q)_{\infty}\,a^{\frac{m}{2}}b^{\frac{n}{2}}}{(q^{1+\frac{n+m}{2}}(ab)^{-1};q)_{\infty}}.
\end{align}

In this letter, we study gamma function limit of identity (\ref{Pentagon-Index}) and present connection to statistical mechanics and superymmetric sphere partition function.  The structure of the paper starts with gamma function limit of three dimensional index in section \ref{2}. Afterwards, we study different gamma function limits of $S^3_b$ partition function in section \ref{3}.

\section{Gamma function limit}\label{2}
%%%%%%%%%%%%%%%%%%%%%%%%%%%%%%%%%%%
In this section we consider the gamma function limit of the pentagon identity related to the three dimensional index. This limit corresponds to the new pentagon identities in terms of gamma function. This pentagon identity have interpretation as the identification of the sphere partition function for the dual  theories.

In order to consider the gamma function limit we take the following limit 
\begin{equation}\label{Limit}
        \lim_{q \rightarrow 1} \frac {(q^\alpha;q)_\infty}{(q^\beta;q)_\infty} (1-q)^{\alpha-\beta}= \frac{ \Gamma (\beta)}{ \Gamma (\alpha)},
\end{equation}
from the pentagon identity (\ref{Pentagon-Index}) which gets the following form
\begin{align}\label{qSeib} \nonumber
     \sum_{m=-\infty}^{\infty}  \oint \prod_{i=1}^3 \frac{(q^{1+(m+n_i)/2}/a_iz,q^{1+(n_i-m)/2} z/b_i ;q)_\infty}{(q^{(m+n_i)/2}a_iz,q^{(n_i-m)/2} b_i/z ;q)_\infty} \frac{1}{z^{3m}} \frac{dz}{2\pi i z} \\ =\frac {2}{ \prod_{i=1}^3 a_i^{m_i}b_i^{n_i}}  \prod_{i,j=1}^{3} \frac {(q^{1+(m_i+n_j)/2}/a_i b_j;q)_\infty}{(q^{(m_i+n_j)/2}a_i b_j;q)_\infty}, 
\end{align}
with the conditions $\prod_{i=1}^3a_i=\prod_{i=1}^3b_i=q^{\frac{1}{2}}$ and $\sum_{i=1}^{3} m_i=\sum_{i=1}^{3} n_i=0$.  After the limit we obtain the following identity  
\begin{align}\label{Penta-Sphere}\nonumber
  \sum_{m\in \mathcal{Z}}\int_{-\infty}^{\infty}\frac{du}{2\pi }  \prod_{i=1}^{3} \frac{\Gamma(\frac{m+n_i}{2}+\alpha_i+ iu)}{\Gamma(1+\frac{ m+n_i}{2}-\alpha_i- iu)}&\frac{\Gamma(\frac{-m+m_i}{2}+\beta_i- iu)}{\Gamma(1+\frac{ -m+m_i}{2}-\beta_i+ iu)} \\\qquad&=\prod_{i,j=1}^{3} \frac{\Gamma(\alpha_i+\beta_j+\frac{n_i+m_j}{2})}{\Gamma(1-\alpha_i-\beta_j-\frac{n_i+m_j}{2})},
\end{align}
with the conditions $\sum_{i=1}^3\alpha_i=\sum_{i=1}^3\beta_i=\frac{1}{2}$ and $\sum_{i=1}^3m_i=\sum_{i=1}^3n_i=0$.
It is interesting that this identity corresponds to the identification of the sphere partition functions \cite{Benini:2016qnm} for the three dimensional duality which we discussed below the formula (\ref{B_{2,2}}). It would be interesting to study such duality in two dimension.

Identity (\ref{Penta-Sphere}) is equivalent to the following pentagon identity
\begin{align}\nonumber
    \sum_{m\in \mathcal{Z}}&\int_{-\infty}^{\infty}\frac{du}{2\pi }  \prod_{i=1}^{3}\mathcal{B}\big(\alpha_i+iu,n_i+m;\beta_i-iu,m_i-m\big)\\
    &\quad=\mathcal{B}\big(\alpha_1+\beta_2,n_1+m_2;\alpha_3+\beta_1,n_3+m_1\big)\mathcal{B}\big(\alpha_2+\beta_1,n_2+m_1;\alpha_3+\beta_2,n_3+m_2\big),
\end{align}
where $\mathcal{B}(a,n;b,m)$ reads 
\begin{align}\label{Gamma2}
    \mathcal{B}(a,n;b,m):=\frac{\Gamma(a+\frac{n}{2})}{\Gamma(1-a+\frac{n}{2})}\frac{\Gamma(b+\frac{m}{2})}{\Gamma(1-b+\frac{m}{2})}\frac{\Gamma(1-a-b+\frac{n+m}{2})}{\Gamma(a+b+\frac{n+m}{2})}.
\end{align}
 Statistical mechanics interpretation of  this pentagon identity is equivalent to the star-triangle relation with continuous and discrete spin parameters.\footnote{See \cite{Spiridonov:2010em, Gahramanov:2015cva, Gahramanov:2017ysd} for the relation between supersymmetric dualities and star-triangle relation. More information about pentagon identities and star-triangle relations see \cite{Bozkurt:2019xno}.} 

\section{Other Gamma Function Limits}\label{3}
%%%%%%%%%%%%%%%%%%%%%%%%%%%%%%%%%%%
In this section we will consider other gamma function solutions of pentagon identity. There are two different gamma function limits of $S^3_b$ partition function for dual theories.

If we consider the following limit 
\begin{align}\label{limit}
\lim_{\omega_2\rightarrow \infty}    \gamma^{(2)}(z;\omega_1,\omega_2)=\big(\frac{\omega_2}{2\pi \omega_1}\big)^{\frac{1}{2}-\frac{z}{\omega_1}}\frac{\Gamma(z/\omega_1)}{2\pi},
\end{align}
for the pentagon identity (\ref{Pentagon-Hyperbolic}), we will get the pentagon identity in terms of gamma function  \cite{Kashaev:2012cz}.
Pentagon identity (\ref{PI}) gets the following form with  $\mathcal{B}(x,y):=\frac{\Gamma(x)\Gamma(y)}{\Gamma(x+y)}$ \cite{Kashaev:2015tma} 
\begin{align}\label{Gamma1}
 \int_{-i\infty}^{i\infty} \frac{du}{2\pi i}  \prod_{i=1}^{3} \mathcal{B}\big(a_i+u,b_i-u\big)=\mathcal{B}\big(a_1+b_2,a_3+b_1\big)\mathcal{B}\big(a_2+b_1,a_3+b_2\big).
\end{align}

From statistical mechanics point of view identity (\ref{Pentagon-Hyperbolic}) is equivalent to the  star-triangle relation \cite{Spiridonov:2010em} for Faddeev-Volkov model \cite{Faddeev:1993pe} with continuous spin. In \cite{Bazhanov:2007vg}, authors also studied one of the main physical regime so-called ``strongly coupled'' regime of Faddeev-Volkov model corresponds to $\omega_1\rightarrow 1$. We note that in that case star-triangle relation is equivalent to the identity (\ref{Penta-Sphere}) obtained in this paper. 

\section{Conclusion}
%%%%%%%%%%%%%%%%%%%%%%%%%%%%%%%%%%%
We study different gamma function limits of the pentagon identities related to three-dimensional index and $S^3_b$ partition function. We obtain a new solution to the pentagon identity in terms of gamma function (\ref{Gamma2}) by considering $q\rightarrow 1$ limit of the index. This solution is different from the one in literature (\ref{Gamma1}). Pentagon identity  (\ref{Penta-Sphere}) is equivalent to the ``strongly coupled'' regime of Faddeev-Volkov model. We conclude that $|b|\rightarrow 1$ limit of $S^3_b$ partition function for the duality is equivalent to  $q\rightarrow 1$ of three dimensional index for the same duality.

\section*{Acknowledgements}
I am indebted to Ilmar Gahramanov for reading the manuscript and fruitful discussions. I am thankful to Elmar Asgerov and Altar \c{C}i\c{c}eksiz for their contributions in the early stages of this work. I am specially grateful to Do\~{g}u D\"{o}nmez and Deniz Bozkurt for the discussion about the topic. I am also thankful to the Simons Center for Geometry and Physics for warm hospitality and support where the main part of the project has done during the program ``Exactly Solvable Models of Quantum Field Theory and Statistical Mechanics'' September 4 -- November 30, 2018. This project is partially supported by TUBITAK under Project No. 117F327 and by BAP Project (no. 2018-18), funded by Mimar Sinan Fine Arts University, Istanbul,
Turkey.\\

\end{document}